\documentclass[fleqn,usenatbib]{mnras}

\usepackage{newtxtext,newtxmath}

\usepackage[T1]{fontenc}
\usepackage{ae,aecompl}

\usepackage{graphicx}	
\usepackage{amsmath}	

\title[vAPP coronagraphic observations of HR~2562~B]{High-contrast observations of brown dwarf companion HR~2562~B with the vector Apodizing Phase Plate coronagraph}

\author[B. Sutlieff et al.]{Ben J. Sutlieff,$^{1,2}$\thanks{E-mail: b.j.sutlieff@uva.nl}
Alexander J. Bohn,$^{2}$
Jayne L. Birkby,$^{3,1,4}$
Matthew A. Kenworthy,$^{2}$
\newauthor Katie M. Morzinski,$^{5}$
David S. Doelman,$^{2}$
Jared R. Males,$^{5}$
Frans Snik,$^{2}$
Laird M. Close,$^{5}$
\newauthor Philip M. Hinz,$^{6}$
and David Charbonneau$^{4}$
\\\\
$^{1}$Anton Pannekoek Institute for Astronomy, University of Amsterdam, Science Park 904, 1098 XH Amsterdam, The Netherlands\\
$^{2}$Leiden Observatory, Leiden University, P.O. Box 9513, 2300 RA Leiden, The Netherlands\\
$^{3}$Department of Astrophysics, University of Oxford, Denys Wilkinson Building, Keble Road, Oxford, OX1 3RH, United Kingdom\\
$^{4}$Center for Astrophysics \textbar~Harvard \& Smithsonian, 60 Garden Street, Cambridge, MA 02138, USA\\
$^{5}$Steward Observatory, University of Arizona, 933 N. Cherry Ave., Tucson, AZ 85721, USA\\
$^{6}$Department of Astronomy and Astrophysics, University of California, Santa Cruz, 1156 High St, Santa Cruz, CA 95064, USA
}

\date{Accepted XXX. Received YYY; in original form ZZZ}

\pubyear{2020}

\begin{document}
\label{firstpage}
\pagerange{\pageref{firstpage}--\pageref{lastpage}}
\maketitle

\begin{abstract}
The vector Apodizing Phase Plate (vAPP) is a class of pupil plane coronagraph that enables high-contrast imaging by modifying the Point Spread Function (PSF) to create a dark hole of deep flux suppression adjacent to the PSF core. Here, we recover the known brown dwarf HR~2562~B using a vAPP coronagraph, in conjunction with the Magellan Adaptive Optics (MagAO) system, at a signal-to-noise of S/N = 3.04 in the lesser studied L-band regime. The data contained a mix of field and pupil-stabilised observations, hence we explored three different processing techniques to extract the companion, including Flipped Differential Imaging (FDI), a newly devised Principal Component Analysis (PCA)-based method for vAPP data. Despite the partial field-stabilisation, the companion is recovered sufficiently to measure a 3.94~\textmu m narrow-band contrast of ($3.05\pm1.00$)~$\times$~10$^{-4}$ ($\Delta$m\textsubscript{3.94\textmu m}~=~8.79$\pm$0.36~mag). Combined with archival GPI and SPHERE observations, our atmospheric modelling indicates a spectral type at the L/T transition with mass M~=~29$\pm$15~M\textsubscript{Jup}, consistent with literature results. However, effective temperature and surface gravity vary significantly depending on the wavebands considered (1200$\leq$T\textsubscript{eff}(K)$\leq$1700 and 4.0$\leq$log(g)(dex)$\leq$5.0), reflecting the challenges of modelling objects at the L/T transition. Observations between 2.4-3.2~\textmu m will be more effective in distinguishing cooler brown dwarfs due to the onset of absorption bands in this region. We explain that instrumental scattered light and wind-driven halo can be detrimental to FDI+PCA and thus must be sufficiently mitigated to use this processing technique. We thus demonstrate the potential of vAPP coronagraphs in the characterisation of high-contrast substellar companions, even in sub-optimal conditions, and provide new, complementary photometry of HR~2562~B.

\end{abstract}




\section{Introduction}
\begin{figure*}
	\centering
      \textcolor{white}{\frame{\includegraphics[]{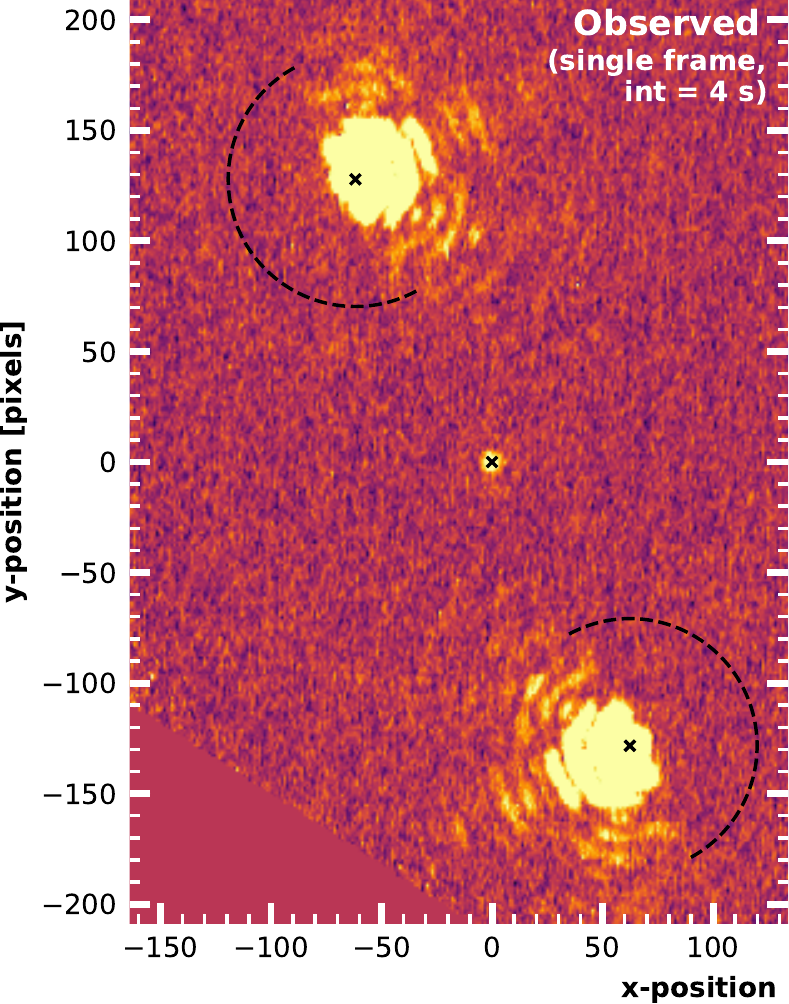}}}%
      \textcolor{white}{\frame{\includegraphics[]{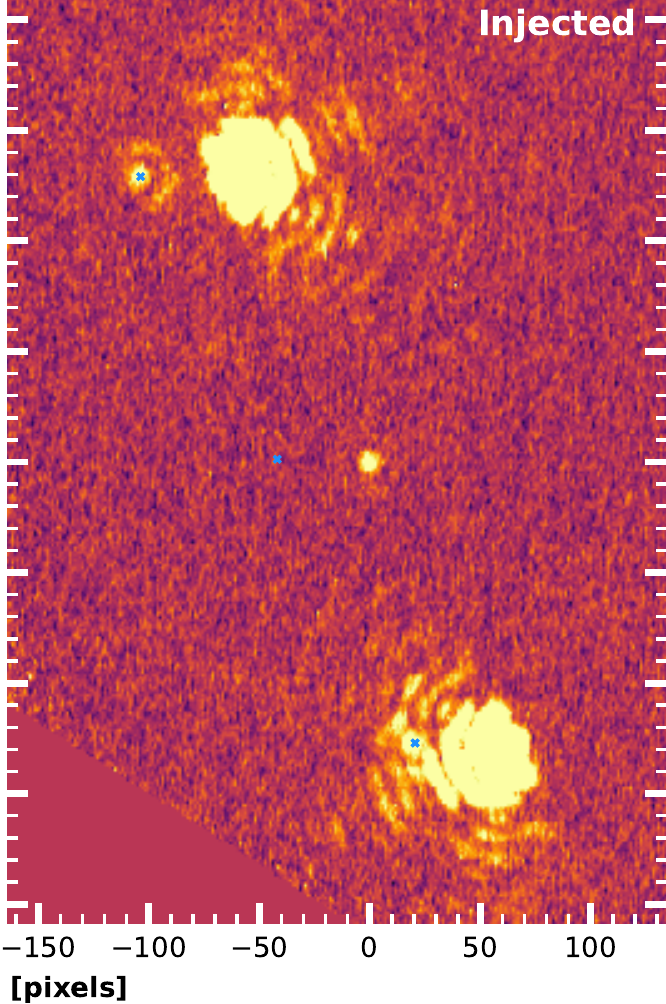}}}%
    \caption{HR~2562 as it appears in a single science frame from the MagAO vAPP coronagraph after pre-processing (left). The three PSFs characteristic of the vAPP are visible, with the centres of the PSF cores indicated by black crosses. At the top and bottom of the image are the coronagraphic PSFs with complementary D-shaped dark holes of deep stellar flux suppression (bounded by black dashed arcs), and the unmodified leakage PSF appears at the origin. The spatial scale shows the differential offsets (in pixels) of the coronagraphic PSFs on the detector with respect to the leakage term, however all three PSFs represent the same position on sky. On the right is the same frame with an artificial companion injected at a contrast of $4.0$~$\times$~10$^{-2}$ ($\Delta$m\textsubscript{3.94\textmu m}~=~3.5~mag) and separation of 41~pixels. The companion PSFs (indicated by blue crosses) have the same shape and structure as the three stellar PSFs. The injected companion can therefore be seen both in the dark hole of the top coronagraphic stellar PSF and, when compared to the left panel, obscured by the flux of the bottom one. The leakage term corresponding to the companion is also present to the left of the stellar leakage term, but is too faint to be visible. Both images are presented with an arbitrary logarithmic colour scale. The frame is not aligned to north, and the lower left corner is masked due to bad pixels.}
    \label{fig:vapp_psf}
\end{figure*}

The detection and characterisation of planetary-mass and brown dwarf substellar companions through high-contrast imaging is reliant on coronagraphs that suppress the diffraction haloes of their host stars. A combination of innovative coronagraph design and optimal post-processing strategy is required to achieve deep contrast ratios at the smallest angular separations currently accessible to ground-based astronomy, where the companion flux can be dominated by quasistatic speckles of residual starlight \citep{1999PASP..111..587R, 2007ApJ...654..633H, 2013A&A...554A..41M}. The ever-growing sample of imaged planetary-mass \citep[e.g.][]{2008Sci...322.1348M, 2010Sci...329...57L, 2015Sci...350...64M, 2017A&A...605L...9C, 2018A&A...617A..44K, 2019NatAs...3..749H, 2019A&A...626A..99J, 2020ApJ...898L..16B} and brown dwarf \citep[e.g.][]{2005A&A...430.1027C, 2015ApJ...805L..10H, 2015ApJ...811..103M, 2016A&A...593A.119M, 2019A&A...626A..99J, 2020ApJ...902L...6W, 2020ApJ...904L..25C} companions highlights the success of the technique. However, many of the instruments involved in these discoveries use focal-plane coronagraphs \citep{2005ApJ...618L.161S, 2012SPIE.8442E..04M, 2018SPIE10698E..2SR} which are inherently susceptible to tip/tilt instabilities, primarily resulting from telescope vibrations, that limit their ability to reach deeper contrast ratios \citep{2014SPIE.9148E..1UF, 2017ApJ...834..175O}. Conversely, vector Apodizing Phase Plate (vAPP) coronagraphs reside in the pupil plane and are therefore inherently insensitive to these tip/tilt instabilities. This intrinsic stability also facilitates beam-switching, which is advantageous in the thermal infrared for the removal of background flux. By adjusting the phase of the incoming wavefront, the vAPP modifies the Point Spread Functions (PSFs) of all objects in the field of view to create a `dark hole', a region of deep flux suppression, adjacent to the PSF core \citep{2014OExpr..2230287O, 2017SPIE10400E..0UD, 2017SPIE10400E..0VP, 2020SPIE11448E..3WB}. The 6.5-m Magellan Clay telescope at Las Campanas Observatory (LCO) hosts a vAPP coronagraph for use in combination with the Magellan Adaptive Optics (MagAO) system \citep[][]{2012SPIE.8447E..0XC, 2014SPIE.9148E..04M}. This vAPP \citep[described by][]{2017ApJ...834..175O} uses a polarization grating to split incoming light according to its circular polarization, resulting in two complementary coronagraphic PSFs each with a 180\textdegree{} D-shaped dark hole on the opposing side, enabling a full view of the region around a target star in a single image \citep{2012SPIE.8450E..0MS, 10.1117/12.2056096}. The size of these dark holes is wavelength dependent, with inner and outer working angles of 2~-~7~$\lambda$/D. A faint and unmodified `leakage' PSF also appears halfway between the two coronagraphic PSFs. These three PSFs are shown in Figure~\ref{fig:vapp_psf}, with the centres of the PSF cores indicated by black crosses. The centres of these PSFs were found by fitting the PSF core with a 2D Gaussian and identifying the location of the peak flux. The leakage term collates the polarization leakage (i.e. the small fraction of light that does not receive the phase adjustment, \citealt{2020PASP..132d5002D}), and can be useful for photometric monitoring of companions or other objects detected in the dark hole (Sutlieff et al., in prep.), depending on the phase design of the vAPP in question. The deep speckle suppression is highly advantageous, but comes at the expense of a few factors. For example, a companion will only be visible in the dark hole of one coronagraphic PSF, hence a loss of overall companion flux of $\sim$50\% \citep{2020PASP..132d5002D}. Further, due to the use of a polarization grating to split the coronagraphic PSFs, their separation is wavelength-dependent and all three PSFs are laterally smeared across the detector \citep{2017ApJ...834..175O}. However, narrow-band filters with a full width at half maximum (FWHM) of $\frac{\Delta\lambda}{\lambda}\leq0.06$ can limit the smearing to $<1\lambda$/D, albeit at the expense of a lower total flux compared to when broad-band filters are used. The deep flux suppression of the vAPP can be further augmented by bespoke data reduction and post-processing strategies designed to remove residual speckles while handling the unique PSF shape, achieving optimal sensitivity to substellar companions in the dark hole. To date, the vAPP at the Large Binocular Telescope has been used to image a protoplanetary disc \citep{2020AJ....159..252W}, and Apodizing Phase Plate coronagraphs (APPs; the predecessor technology to the vAPP, \citealt{2006SPIE.6269E..1NC, 2007ApJ...660..762K}) were successfully used to detect substellar companions at high contrasts \citep{2015MNRAS.453.2378M, 2015MNRAS.453.2533M, 2010ApJ...722L..49Q, 2015ApJ...807...64Q}. However, observations of substellar companions using vAPPs have yet to be reported.

HR~2562 (HD~50571; HIP32775) is an F5V star with an estimated mass of 1.368$\pm$0.018 M$_{\odot}$ \citep{2018A&A...612A..92M} at a distance of 34.007$\pm$0.048 pc \citep{2018A&A...616A...1G, 2018AJ....156...58B}. The key properties of the star are summarised in Table \ref{table:stellar_parameters}. As is common for F-type stars without known membership of a moving group or cluster, the age of the system is not well constrained, with the strongest constraints on the age (450{\raisebox{0.5ex}{\tiny$\substack{+300 \\ -250}$}} Myr) arriving from measurements of the stellar lithium-temperature relationship \citep{2018A&A...612A..92M}. HR~2562 has a circumstellar debris disc at an inclination of 78.0$\pm$6.3\textdegree and position angle of 120.1$\pm$3.2\textdegree, with an inner radius of 38$\pm$20 au and an outer radius of 187$\pm$20 au \citep{2006ApJ...644..525M, 2015MNRAS.447..577M}. Using the Gemini Planet Imager \citep[GPI,][]{2014PNAS..11112661M} in the J-, H-, and K-band \citet{2016ApJ...829L...4K} identified a 30$\pm$15~M\textsubscript{Jup} substellar companion to HR~2562, with an estimated spectral type of L7$\pm$3 at a projected separation of 20.3$\pm$0.3~au (0.618$\pm$0.003$\arcsec$), orbiting coplanar to the debris disc and within the inner gap of the disc. This companion is one of only two detected brown dwarfs orbiting interior to its host debris disc, alongside HD~206893~B \citep{2017A&A...597L...2M}. \citet{2018A&A...612A..92M} and \citet{2018A&A...615A.177M} conducted a further study of the system with the Spectro-Polarimetic High-contrast imager for Exoplanets REsearch \citep[SPHERE,][]{2019A&A...631A.155B} instrument at the Very Large Telescope (VLT), completing an extensive spectrophotometric and astrometric characterisation of the companion through spectral observations in the Y- to J- band range plus broad-band imaging in the H-band. They derive a similar mass of 32$\pm$14~M\textsubscript{Jup} but an early T spectral type. HR~2562 is an ideal target for the MagAO vAPP as the companion separation is at the centre of the dark hole of the vAPP at 3.94 \textmu m (which covers a working angle of 261 - 912 mas at this wavelength) at an achievable contrast \citep[$\Delta$K2~=~$\sim$10.4 mag,][]{2016ApJ...829L...4K}. It is therefore optimal for developing and testing procedures for data reduction and post-processing. Furthermore, photometry of the companion at a wavelength longer than those in previous studies can further constrain physical properties of HR~2562~B, such as effective temperature and surface gravity, and help to resolve the tension in its spectral classification.

In this paper we present the first reported images of a substellar companion using a vAPP coronagraph. In Section \ref{obs} of this paper we describe the observations performed on HR~2562, and in Section \ref{data_red} we outline the data reduction and new post-processing methodology we developed for data obtained with a vAPP. In Section \ref{results} we explain how we obtained our photometric measurements, and fit spectral models and empirical templates to the data to obtain values for the physical parameters of the companion. We then discuss these results in Section \ref{discussion}, and compare them to previous results from the literature. We also discuss the effectiveness and limitations of our post-processing strategy. The conclusions of the paper are presented in Section \ref{conclusions}.
\begin{flushleft}
\begin{table}
\caption{Properties of host star HR~2562.}
\begin{tabular}{p{0.41\columnwidth}p{0.2\columnwidth}p{0.2\columnwidth}}
\hline
Parameter&Value&Reference(s)\\
\hline
Spectral Type&F5V&(1)\\
Right Acension (J2000)&06:50:01.02&(2)\\
Declination (J2000)&-60:14:56.92&(2)\\
Age (Myr)&450{\raisebox{0.5ex}{\tiny$\substack{+300 \\ -250}$}}&(3)\\
Parallax (mas)&29.3767$\pm$0.0411&(2)\\
Distance (pc)&34.007$\pm$0.048&(2, 4)\\
Proper motion (RA, mas yr$^{-1}$)&4.663$\pm$0.084&(2)\\
Proper motion (Dec, mas yr$^{-1}$)&108.377$\pm$0.089&(2)\\
Mass (M$_{\odot}$)&1.368$\pm$0.018&(3)\\
Radius (R$_{\odot}$)&1.334$\pm$0.027&(3)\\
T\textsubscript{eff} (K)&6597$\pm$81&(5)\\
log(g) (dex)&4.3$\pm$0.2&(3)\\
{[Fe/H]}&0.10$\pm$0.06&(3)\\
V (mag)&6.098$\pm$0.010&(6)\\
G (mag)&5.9887$\pm$0.0005&(2)\\
J (mag)&5.305$\pm$0.020&(7)\\
H (mag)&5.128$\pm$0.029&(7)\\
K (mag)&5.020$\pm$0.016&(7)\\
\hline

\end{tabular}
\textbf{References:} (1) \citet{2006AJ....132..161G}; (2) \citet{2018A&A...616A...1G}; (3) \citet{2018A&A...612A..92M}; (4) \citet{2018AJ....156...58B}; (5) \citet{2011A&A...530A.138C}; (6) \citet{2000A&A...355L..27H}; (7) 2MASS \citep{2003yCat.2246....0C}
\label{table:stellar_parameters}
\end{table}
\end{flushleft}
\section{Observations}\label{obs}
We observed the star HR~2562 and its substellar companion \citep[separated by 643.8$\pm$3.2 mas,][]{2018A&A...615A.177M} on the nights of 2017 February 06 (02:47:39 - 05:16:11 UT) and 2017 February 07 (02:08:32 - 07:34:34 UT), with the vAPP coronagraph and the MagAO \citep[][]{2012SPIE.8447E..0XC, 2014SPIE.9148E..04M} system on the 6.5-m Magellan Clay telescope at LCO, Chile. We used the Clio2 Narrow near-IR camera, which has a plate scale of 15.85 mas pixel$^{-1}$ and an array of 1024 x 512 pixels, giving a field of view of 16$\arcsec$x 8$\arcsec$ \citep{2006SPIE.6269E..0US, 2015ApJ...815..108M}. The vAPP was positioned in the pupil stop wheel of Clio2 as described in \citet{2017ApJ...834..175O}, such that three PSFs of the star appeared in a sequence across the short axis of the detector (as shown in Figure~\ref{fig:vapp_psf}), leaving significant room on the long axis for background subtraction by nodding. We used a $\lambda=$3.94 \textmu m narrow-band filter with a width of 90~nm for these observations, which placed the companion at the centre of the dark hole of the top coronagraphic PSF. With this filter, $\frac{\Delta\lambda}{\lambda}=0.023$, so wavelength-dependent radial smearing is limited to $<0.4\lambda$/D. Furthermore, the MagAO system achieves a high Strehl ratio (>90\%) at this wavelength \citep[][]{2017ApJ...834..175O}. Atmospheric conditions were clear throughout the observations. On the first night, seeing was measured at 0.6$\arcsec$ at the beginning of observations. At the start of the second night seeing was poor (1.3$\arcsec$) with no wind, and improved to 0.5-0.6$\arcsec$ seeing by midnight, but with $\sim$13 m~s$^{-1}$~winds. Observations were obtained in a continuous sequence on each night (interrupted only when the adaptive optics loop opened). We obtained 362 and 403 data cubes on the first and second nights, respectively. Each cube contains 10 sub-frames, where each sub-frame represents an integration time of 2~s on the first night and 4~s on the second. The total on-target integration time across both nights is thereby (362 $\times$ 10 $\times$ 2 + 403 $\times$ 10 $\times$ 4) = 23360~s ($\sim$6.5~h). The increased exposure time for the second night was chosen as a compromise to minimize the effect of readout noise without obtaining excessive flux due to the high sky background at 3.94~\textmu m. For background subtraction, we used an ABBA nodding pattern. Dark frames were also obtained at the corresponding exposure times for the science frames at the end of the night. The majority of the data was obtained in field-stabilised mode with the derotator switched on and the companion position fixed in the dark hole. Although this is non-standard for high-contrast imaging, our original intention for these observations was to characterise the stability of the MagAO vAPP over time by identifying fluctuations that correspond to instrumental systematics, hence we wanted to keep souces stationary on the same pixels (Sutlieff et al., in prep.). However, the derotator malfunctioned part way through each night (at 05:01:08 UT on the first night, and 04:44:34 UT on the second), causing the field to rotate during the remainder of the observing sequence. The field rotation when the derotator was off was 4.36\textdegree{}  and 42.29\textdegree{}  on the first and second nights, respectively. This mix of field-stabilised and pupil-stabilised data is not the most optimal approach for high-contrast imaging. Nonetheless, in the latter case, the high field rotation was sufficient enough that we were able to use the Angular Differential Imaging \citep[ADI,][]{2006ApJ...641..556M} technique to reduce quasistatic speckle noise in the data from the second night (as discussed in Section \ref{postprocess}), and determine a flux for the companion in the L-band regime for the first time. All three of the PSFs remained unsaturated in the core. By coincidence, HR~2562 was also observed with SPHERE on the night of 2017 February 07, the second night of our observations \citep{2018A&A...612A..92M, 2018A&A...615A.177M}, providing an exact known position of the companion in our observations.
\begin{figure*}
\centering
  \textcolor{white}{\frame{\includegraphics[scale=0.5]{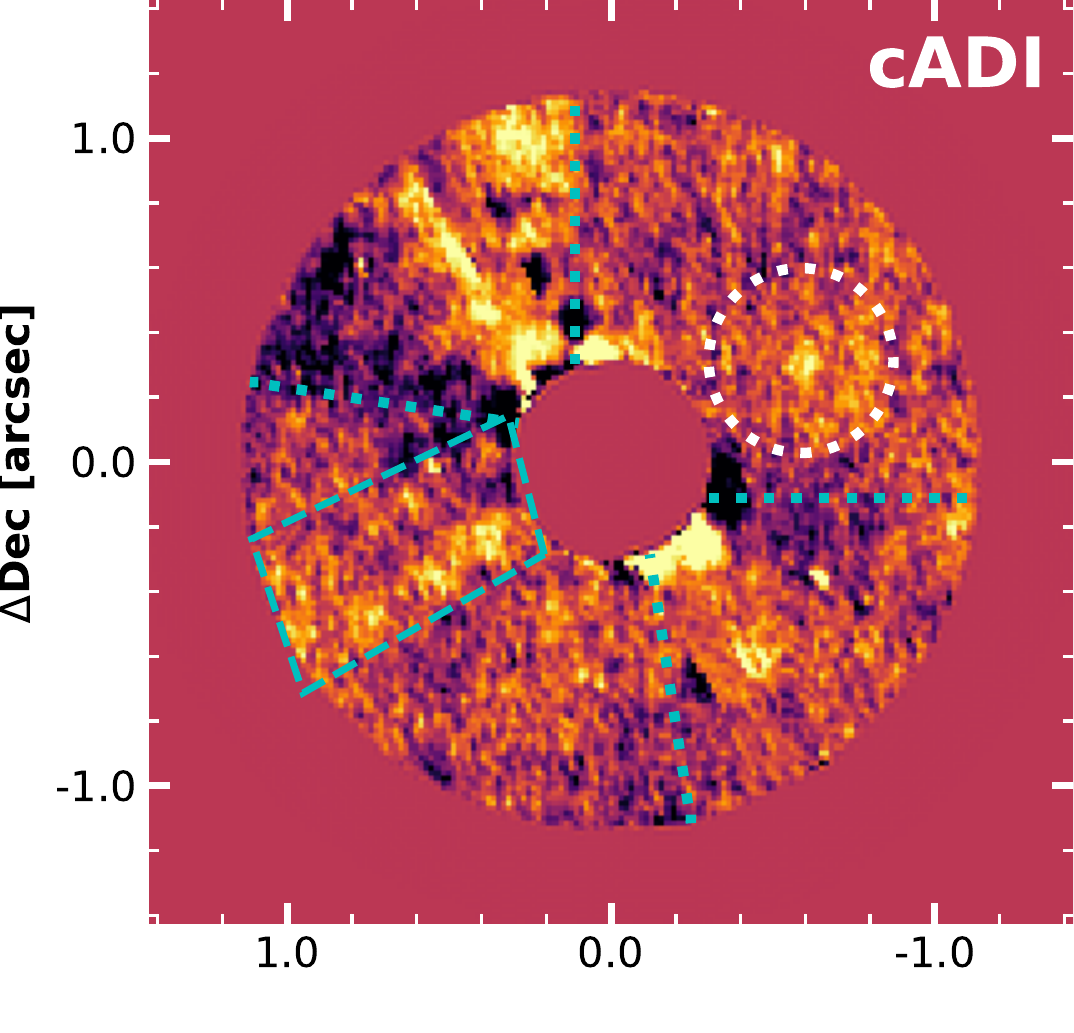}}}%
  \textcolor{white}{\frame{\includegraphics[scale=0.5]{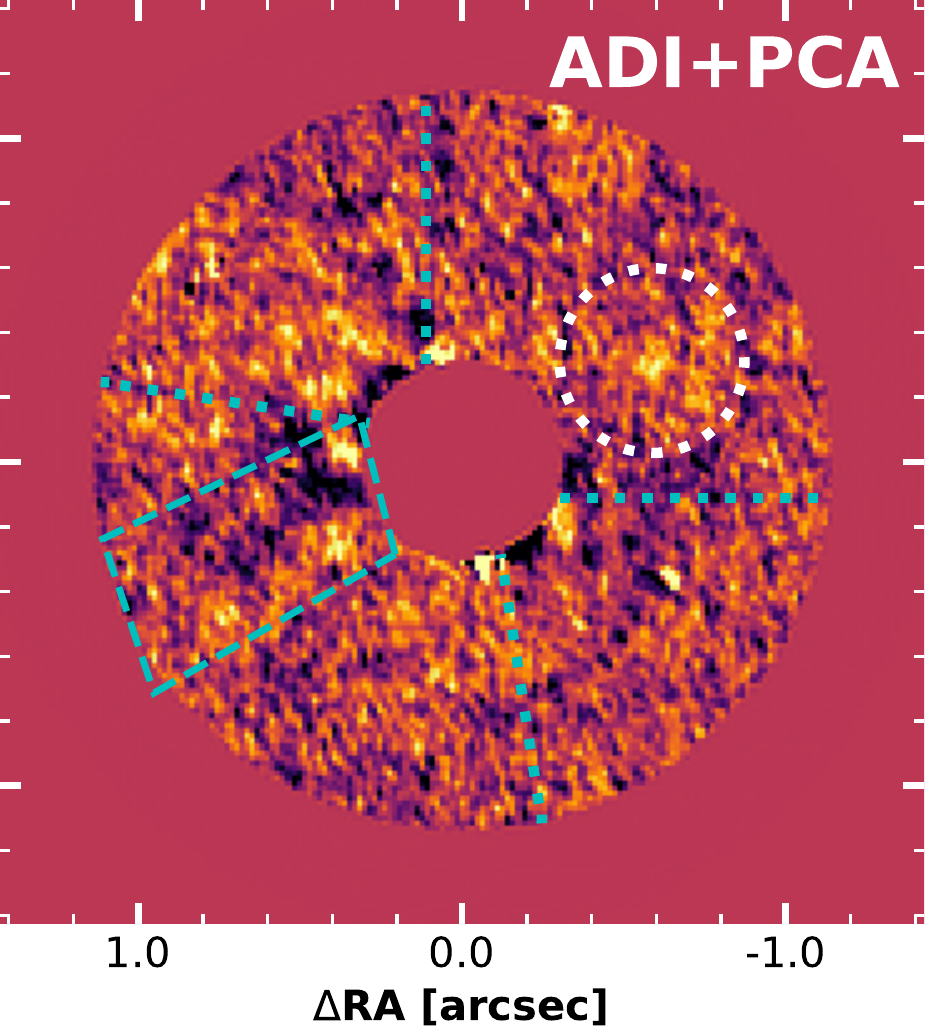}}}%
  \textcolor{white}{\frame{\includegraphics[scale=0.5]{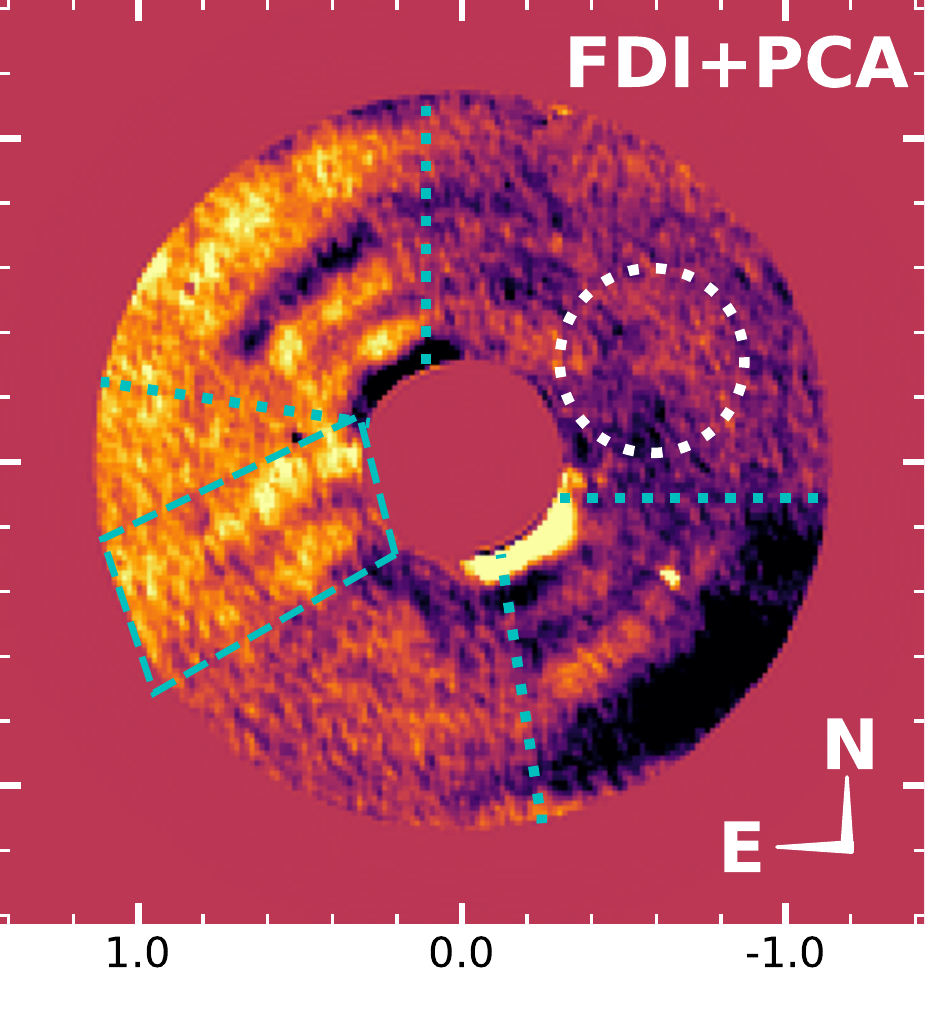}}}%
\caption{The final post-processed images obtained with MagAO+vAPP, as processed using the cADI (left), ADI+PCA (middle), and FDI+PCA (right) algorithms, and covering a total integration time of 14,160 seconds. The two D-shaped dark holes have been stitched together around their common centre. Residual contamination is visible where the dark holes were joined together in the form of bright and dark regions in the north-east and south-west (segments indicated by blue dotted lines). Regions inside the inner working angle of the vAPP and beyond the outer expanse of the vAPP PSF have been masked. The expected location of the companion from concurrent SPHERE observations \citep{2018A&A...615A.177M} is indicated by a dotted white circle in each image. HR~2562~B is detected in the cADI and ADI+PCA images, but not in the FDI+PCA image. The non-detection in the FDI+PCA image is due to symmetry-breaking factors such as instrumental ghosts and wind-driven halo, and is discussed in Section \ref{fdi}. The characteristic butterfly pattern of wind-driven halo can be seen in the FDI+PCA image as extended bright and dark patches immediately and diagonally either side of the masked inner region. A bright spike caused by instrumental scattered light is indicated in the blue dashed box. The bright patch in the south-west is a persistent detector defect that was not removed during the data reduction process. All three images use an arbitrary logarithmic colour scale.}
\label{fig:final_images}
\end{figure*}
\section{Data reduction} \label{data_red}
\subsection{Pre-processing}
To handle the unique PSFs of the vAPP images, we used both standard tools in the literature and bespoke techniques. First, we discarded 10 data cubes from the first night and 49 cubes from the second night that were unusable due to the adaptive optics loop opening during detector exposure. We then corrected non-linear pixels and bad pixels using the formulae and maps described in \citet{2015ApJ...815..108M}. The linearity correction is capable of correcting measured counts up to 45,000 data numbers (DN), where counts above 27,000 DN are considered non-linear. On average, $\sim$0.7\% of pixels in each frame were in this non-linearity regime prior to correction and of these, none were present in the vAPP dark holes except for a small cluster of bad pixels in the top dark hole at the `A' nod position, which were later corrected. Although the bad pixel map did not cover all of the bad pixels in our data, most of the remainder did not lie within or close to the vAPP dark hole. A master dark frame was created for each night by median combining five dark frames with the same array size, integration times, and input offset voltage as our data. The master dark frame was then subtracted from every raw science frame. We created an `A' nod position sky flat and a `B' nod position sky flat by median combining all of the dark-subtracted science frames at the opposite nod position. We normalised each of these sky flats by dividing them by the median number of counts in a region of the frame away from the PSFs. These normalised sky flats were then divided out of the dark-subtracted science frames, removing variations caused by the response of the detector and long-term sky structure throughout the observations. After these calibrations, background subtraction was carried out using the data from the opposing nod positions of the ABBA pattern. For each data cube obtained in the A position, we subtracted the corresponding B position data cube obtained closest in time to the A position cube, and vice versa. To remove any residual background offset, we then subtracted the median of a clean region of the data from each frame. A number of instrumental ghosts and other optical effects resulting from internal reflection within the refractive optics of the setup are visible in the data (see Section \ref{fdi}).

\subsection{Post-processing}\label{postprocess}
Additional post-processing of the data is required to further augment the deep flux suppression of the vAPP and achieve the sensitivity needed to detect HR~2562~B. To do this, we used custom modules based on version 0.6.2 of the PynPoint package for high-contrast imaging data \citep{2019A&A...621A..59S}. Firstly, we cropped each of the two coronagraphic PSFs separately and fit their cores with 2D Gaussians to align the data from both nod positions together, making an image cube for each coronagraphic PSF covering the full sequence. This placed the two nod positions at the same location and removed a linear drift in position across the full observing sequence. Regions inside the inner working angle of the vAPP and beyond the outer expanse of the vAPP PSF were then masked and the two opposing dark holes were joined together. At this stage, we separately applied three different post-processing techniques to the joined dark holes, designed to subtract speckle noise and other residual starlight not suppressed by the vAPP, producing three final images.\\
\textbf{Classical ADI (cADI):} The first of these techniques was classical ADI \citep[cADI,][]{2006ApJ...641..556M}. We constructed a reference PSF by taking the median combination of the data. This reference PSF was then subtracted from the data. After subtraction of the reference PSF, we aligned the images to north according to their parallactic angles and median combined them. Unsurprisingly, as cADI is reliant on the field rotation of the observations to prevent the inclusion of flux from the companion in the reference PSF, we do not detect HR~2562~B in the data from the first night. However, in the final cADI image from the second night (which covered significantly more field rotation), the companion is detected at the expected position in the centre of the right-hand (after north alignment) vAPP dark hole and is shown in the left panel of Figure~\ref{fig:final_images}. This is a marginal detection with a signal-to-noise ratio (S/N) of 3.04. Although this is not at the S/N~=~5 level commonly accepted for a detection in a blind search, it is reinforced by its presence at the known position of the companion measured by \citet{2018A&A...615A.177M}, in data obtained on the same night using SPHERE.\\
\textbf{Principal Component Analysis (ADI+PCA):} The second post-processing technique we applied to the joined dark holes was speckle subtraction via Principal Component Analysis \citep[ADI+PCA;][]{2012MNRAS.427..948A, 2012ApJ...755L..28S, 2014ApJ...780...17M}. We used PCA to construct and subtract a reference PSF consisting of 3 principal components, selected as the number that best removed the visible speckle structure and residuals of the vAPP PSF. The residual images were then aligned to north and median combined as above. As above, this technique did not produce a detection in the data from the first night, as the lack of field rotation led to companion self-subtraction. We again marginally detect HR~2562~B in the final image when ADI+PCA was applied to the second night of data, this time with a S/N of 2.38 (centre, Figure~\ref{fig:final_images}).\\
\textbf{Flipped Differential Imaging (FDI+PCA):} The third algorithm we used to construct and subtract a reference PSF was a new technique relying on the symmetry of the coronagraphic vAPP PSFs (hereafter Flipped Differential Imaging, FDI+PCA). With FDI+PCA, the reference PSF to be subtracted from one coronagraphic PSF is produced by applying the PCA algorithm to the opposing coronagraphic PSF after it has been rotated by 180~degrees. This was recommended by \citet{2017ApJ...834..175O} and builds upon a similar approach in the same paper, which uses the opposing vAPP coronagraphic PSF as a reference directly (without applying PCA). It is also similar to the technique used by \citet{2015ApJ...802...12D}, who applied the ADI+PCA concept using a single non-coronagraphic PSF under 180\textdegree{} rotation as a self-reference. As with ADI+PCA, the reference PSF that we created consisted of 3 principal components. We subtracted then north aligned and median combined to produce the final images. In this case, we do not detect HR~2562~B in the images from either night of data. The final FDI+PCA processed image for the second night of data is shown in the right panel of Figure~\ref{fig:final_images}. The symmetry-breaking factors that have affected the performance of the FDI+PCA algorithm, including instrumental ghosts and wind-driven halo, are discussed in Section \ref{fdi}. As FDI+PCA is not inherently reliant on field rotation like cADI and ADI+PCA, in principle we would expect it to be more effective when applied to the first night of data compared to these techniques. However, it was clear from pre-processing that the asymmetric features would have an even stronger effect without field rotation, and that the increased effect of readout noise due to the shorter exposure time on the first night further inhibits detection of the companion.

As we only detect the companion in the data from the second night of observations (which covers a total integration time of 14,160 s), we continue with the data from this night only for the remainder of our analysis. In each of the final images, contamination is seen where the edges of the dark holes were joined together, visible as structured bright and dark patches in the north-east and south-west regions (see segments indicated by blue dotted lines in Figure~\ref{fig:final_images}). However, the region surrounding the expected companion location (based on the concurrent SPHERE observations, \citealt{2018A&A...615A.177M}) is unaffected by this as it is positioned centrally in the vAPP dark hole.
\section{Results}\label{results}
\begin{figure*}
	\includegraphics[scale=0.95]{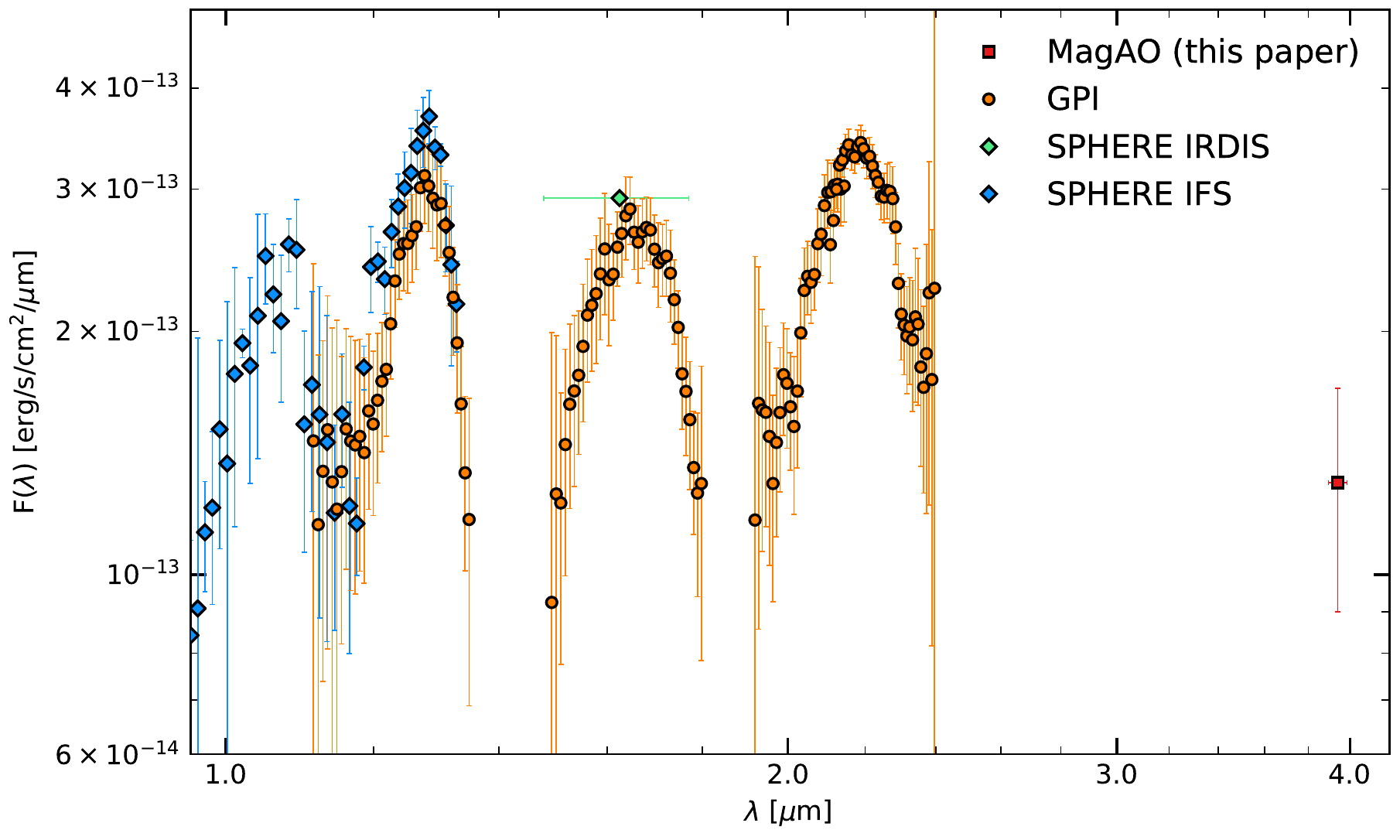}
    \caption{The spectrum of HR~2562~B including our 3.94 \textmu m observation with MagAO+vAPP (red square) alongside all previous photometric data: SPHERE IRDIS photometry using the H broad-band filter (turquoise diamond), SPHERE IFS data in the Y and J bands (blue diamonds), and GPI spectral data in the J, H, K1, and K2 bands (orange circles). The errorbars in the wavelength direction correspond to filter width, or in the case of the IFS and GPI spectral datapoints, Gaussian widths corresponding to the resolution of the respective spectrograph in the relevant band (see \ref{results_models}). The width of the MagAO 3.94 \textmu m narrow-band filter is 90 nm. Some errors are smaller than the symbols.}
    \label{fig:sed_full}
\end{figure*}
\begin{figure*}
	\includegraphics[]{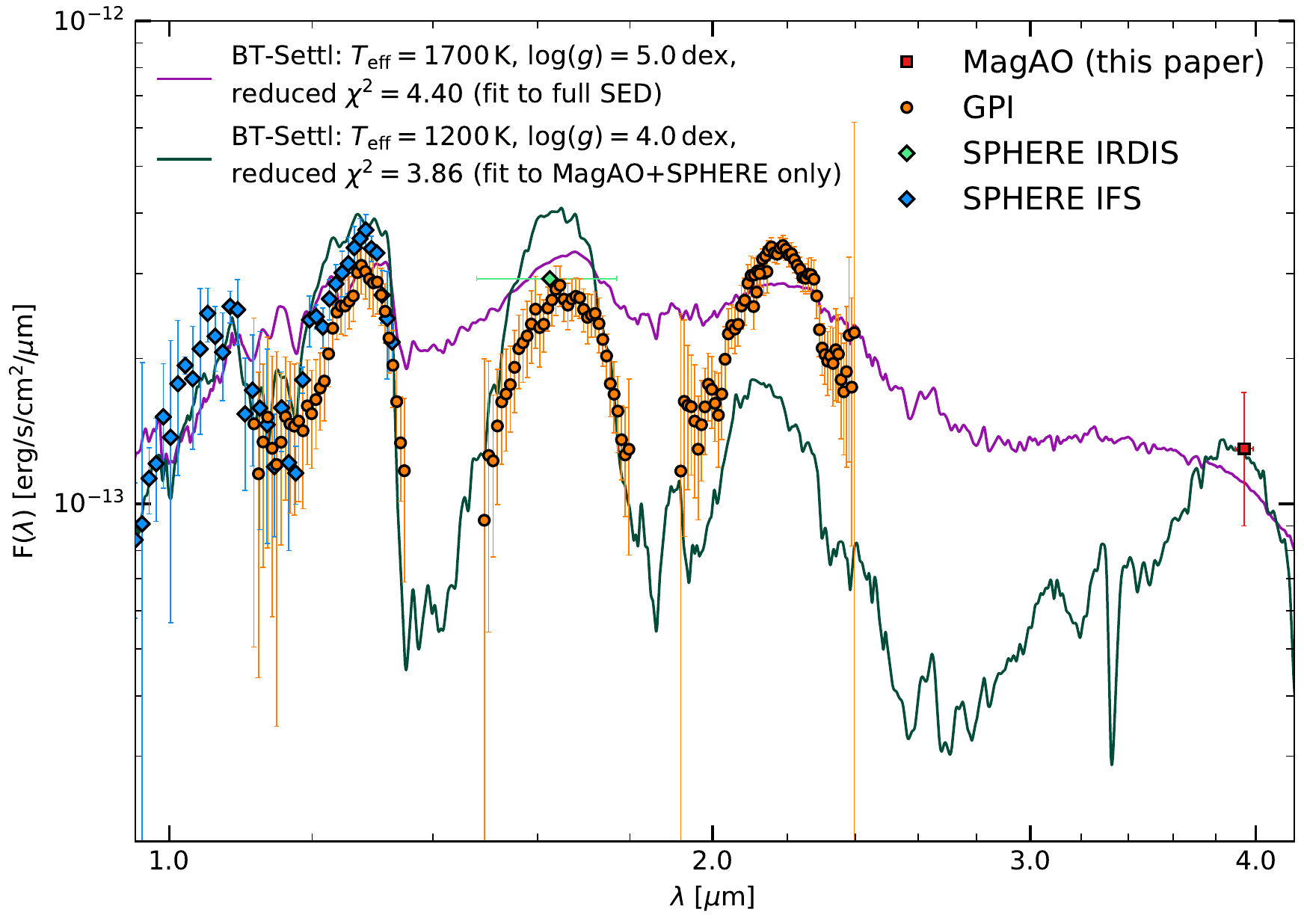}
    \caption{The best fit BT-Settl models to the photometry of HR~2562~B. 
    The purple line shows the best fit to the full SED of HR~2562~B, with T\textsubscript{eff} = 1700 K and log(g) = 5.0 dex, whereas the green line shows the best fit to the MagAO + SPHERE-only subset of data, with T\textsubscript{eff} = 1200 K and log(g) = 4.0 dex. Both models have a metallicity of {[Fe/H]} = 0. Significantly different best fit models are found depending on the wavelength range considered, with very different physical parameters allowed while still providing an equally good fit. Note the large difference in the models between 2.4-3.2 \textmu m.}
    \label{fig:model_multi}
\end{figure*}
The cADI reduction, using observations from the second night only, gives the highest S/N for the companion in the final images (see Figure \ref{fig:final_images}), so we proceed with this technique for the remainder of our analysis, noting that it contains a mix of field- and pupil-stabilised data. The companion was not detected in the first night of data, which was primarily obtained in field-stabilised mode. Either greater photon collecting power or targets with lower contrasts are required to successfully detect companions in field stabilised mode.
\subsection{Photometric measurement}\label{phot_measure}
We measured the contrast ratio of HR~2562~B by injecting scaled negative template companions into the data after pre-processing at the known position of HR~2562~B, following the approach of \citet{2011A&A...528L..15B, 2011ApJ...739L..41G, 2010Sci...329...57L}. The PSFs of companions observed using a vAPP coronagraph have the same shape and structure as the stellar PSFs, i.e., two coronagraphic PSFs and a leakage PSF, all offset from the stellar PSF. However, typically only the coronagraphic PSF in the dark hole is seen, while the other is obscured by the bright coronagraphic stellar PSF, and the companion's leakage PSF is too faint to be detectable (right panel, Figure \ref{fig:vapp_psf}). Template companion injection is therefore only required around the coronagraphic stellar PSF where the companion resides in the dark hole, as only this companion PSF contributes to the detection. We produced this PSF template by median combining the corresponding unsaturated coronagraphic PSF of the star in the pre-processed images and cropping to the first Airy ring. We then scaled the flux of the template relative to the coronagraphic stellar PSF and subtracted it at the location of the companion in the pre-processed data, iterating over different values for the contrast ratio in a grid ranging from contrasts of 8.4$\leq$ $\Delta$m\textsubscript{3.94\textmu m}(mag)$\leq$9.4 with a step size varying from 0.1 to 0.01 as the value was refined. For each injection, we applied cADI as described in Section~\ref{postprocess}. The contrast measurement was then taken as the value which minimized the root mean square in an aperture at the companion location after the negative injection. We also iterated over a grid of possible positions for the companion and found a companion separation of 665.4$\pm$24.0~mas and position angle of 297.3$\pm$2.3\textdegree. These values are consistent with those of \citet{2018A&A...615A.177M} to within 1$\sigma$, who observed HR~2562 with SPHERE on the same night as these observations and found a companion separation of 643.8$\pm$3.2~mas and position angle of 297.51$\pm$0.28\textdegree. The relatively large uncertainties on our position measurements can likely be attributed to the photometric extraction process, which is intrinsically less accurate in the low S/N regime of our measurement. Despite this, the difference between the SPHERE position and our position affects the contrast measurement at the millimagnitude level only. We measure the 3.94~\textmu m contrast to be ($3.05\pm1.00$)~$\times$~10$^{-4}$ ($\Delta$m\textsubscript{3.94\textmu m}~=~8.79$\pm$0.36~mag). We calculated the measurement error on this value following \citet{2015ApJ...815..108M}, which uses the S/N of the companion in the final image. We measured a S/N of 3.04 for the companion by dividing the Gaussian-smoothed peak height of the companion by the standard deviation in an annulus centred on the companion location with inner and outer radii of 1 $\times$ FWHM and 2 $\times$ FWHM wide, respectively. The uncertainty can primarily be attributed to the quasistatic speckle noise throughout the observations. This error bar is relatively large compared to literature measurements of companion contrast, again reflecting the photometric extraction process in the low S/N regime of the detection. The causes of this low S/N are discussed in Section \ref{phot}.\\
The star does not have flux calibrated observations in the 3.94~\textmu m filter. To convert our contrast value to a measurement of the physical flux of the companion, we used the Virtual Observatory SED Analyzer \citep[VOSA,][]{2008A&A...492..277B} to fit the Spectral Energy Distribution (SED) of the host star and calculate the stellar flux at 3.94 \textmu m. We included literature photometry of HR~2562 from Gaia (\citealt{2018A&A...616A...1G}), 2MASS (\citealt{2006AJ....131.1163S}) and WISE (\citealt{2010AJ....140.1868W}) catalogues, and fitted a grid of BT-Settl models \citep{2012RSPTA.370.2765A} using a chi-square test, assuming a distance of 34.01 pc (Gaia DR2) and an extinction of A\textsubscript{V} = 0.07 mag from the extinction map of \citet{2006LNEA....2..189M}. The best fit model had T\textsubscript{eff} = 6600~K, log(g) = 4~dex, and [Fe/H] = 0.5, which are in good agreement with the values derived by \citet{2018A&A...612A..92M}. Evaluating this model in the 3.94 \textmu m filter profile of MagAO/Clio2 and multiplying by our contrast measurement of ($3.05\pm1.00$)~$\times$~10$^{-4}$, we obtain a physical flux of F\textsubscript{3.94\textmu m}~=~($1.3\pm0.4$)~$\times$~10$^{-13}$ erg s$^{-1}$ cm$^{-2}$ micron$^{-1}$ for HR~2562~B at 3.94 \textmu m. This value is shown in Figure~\ref{fig:sed_full} alongside the GPI spectrum from \citet{2016ApJ...829L...4K} in the J, H, K1, and K2 bands; as well as the Y,J SPHERE IFS spectrum and SPHERE IRDIS H-broad-band datapoint from \citet{2018A&A...612A..92M}. The SPHERE IFS and GPI spectra are comparable where they overlap in the J-band, with a small systematic offset within the 1$\sigma$ error bars at $\sim$1.28 \textmu m. \citet{2018A&A...612A..92M} note the possibility of systematic offsets between GPI and SPHERE photometry, likely caused by differences in the algorithms used for processing data, extracting spectra and calibrating the flux \citep{2017AJ....154...10R, 2017A&A...603A..57S}. We nonetheless include the data from both instruments in our analysis of the companion SED, considering theoretical model and empirical template fits to both the entire SED, and subsets that exclude individual instruments (see Sections~\ref{results_models}~-~\ref{results_templates}).

\subsection{Spectral fitting}
\subsubsection{Theoretical atmospheric models}\label{results_models}
\begin{flushleft}
\begin{table*}
\caption{Estimated physical properties of HR~2562~B. The reported errors on the effective temperature and surface gravity are the largest of either the statistical error or the BT-Settl model grid spacing. The errors on the luminosities are the statistical errors. Derived values for the mass of HR~2562~B are found by evaluating the derived values of log(L/L$_\odot$) with BTSettl and AMES-Dusty isochrones across the 200-750 Myr age range of the system, and the corresponding mass ratio with respect to the primary, $q$ (see Section \ref{mass_estimation}). The error on the age of the system dominates the errors on the mass and $q$.}
\begin{tabular}{c|c|c|c|cc|cc}
\hline
Data&T\textsubscript{eff}(K)&log(g) (dex)&log(L/L$_\odot$)& \multicolumn{2}{c|}{Mass (M\textsubscript{Jup})} & \multicolumn{2}{c|}{Mass ratio $q$} \\
 & & & &BT-Settl&AMES-Dusty&BT-Settl&AMES-Dusty\\
\hline
Full SED&1698{\raisebox{0.5ex}{\tiny$\substack{+100 \\ -100}$}}&4.98{\raisebox{0.5ex}{\tiny$\substack{+0.5 \\ -0.5}$}}&-4.60{\raisebox{0.5ex}{\tiny$\substack{+0.03 \\ -0.01}$}}&30.7{\raisebox{0.5ex}{\tiny$\substack{+9.7 \\ -12.1}$}}&33.3{\raisebox{0.5ex}{\tiny$\substack{+10.0 \\ -11.3}$}}&0.021{\raisebox{0.5ex}{\tiny$\substack{+0.007 \\ -0.008}$}}&0.023{\raisebox{0.5ex}{\tiny$\substack{+0.007 \\ -0.008}$}}\\

MagAO+SPHERE&1168{\raisebox{0.5ex}{\tiny$\substack{+132 \\ -100}$}}&4.22{\raisebox{0.5ex}{\tiny$\substack{+0.78 \\ -0.5}$}}&-4.87{\raisebox{0.5ex}{\tiny$\substack{+0.10 \\ -0.11}$}}&25.1{\raisebox{0.5ex}{\tiny$\substack{+9.9 \\ -10.9}$}}&26.3{\raisebox{0.5ex}{\tiny$\substack{+10.7 \\ -12.2}$}}&0.018{\raisebox{0.5ex}{\tiny$\substack{+0.007 \\ -0.008}$}}&0.018{\raisebox{0.5ex}{\tiny$\substack{+0.007 \\ -0.008}$}}\\
\hline

\end{tabular}
\label{table:model_posteriors}
\end{table*}
\end{flushleft}
\begin{figure*}
	\includegraphics[scale=0.95]{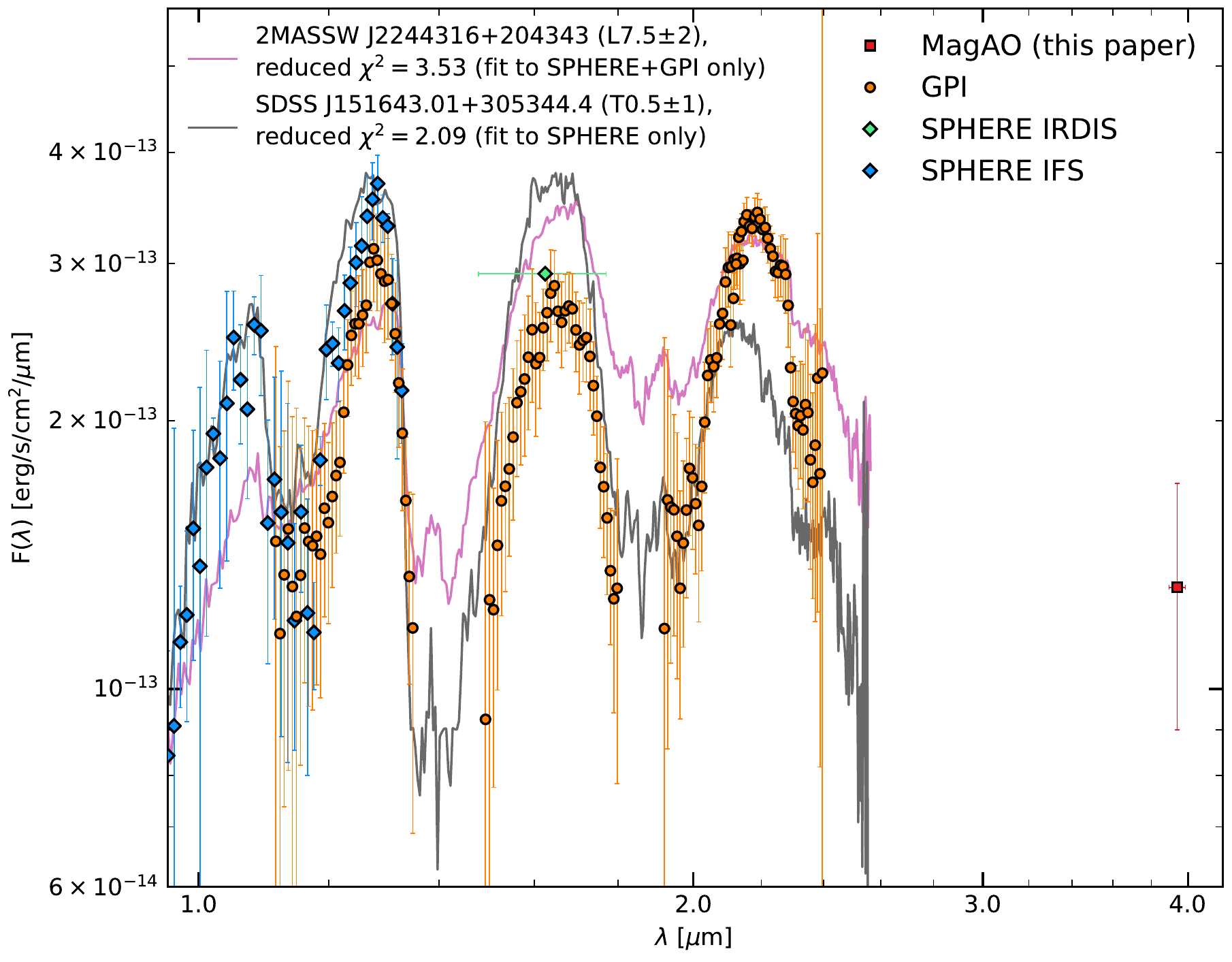}
    \caption{The best fit empirical template spectra to the photometry of HR~2562~B, from a set of L and T dwarf templates taken from the SpeX Prism Spectral Libraries \citep{2014ASInC..11....7B}. The pink line shows the best fit to the combined SPHERE + GPI data, while the grey line shows the best fit model to the SPHERE data only. These templates only extend to $\lambda$ = 2.56 \textmu m, and so do not reach the 3.94 \textmu m wavelength of the MagAO datapoint, which is shown for reference.
    }
    \label{fig:template_multi}
\end{figure*}
To determine the physical properties of HR~2562~B, we followed the approach of \citet{2020MNRAS.492..431B}, using a linear least squares approach to fit grids of theoretical spectra to the photometric data. We selected a grid of BT-Settl models\footnote{Models downloaded from: \url{http://perso.ens-lyon.fr/france.allard/}} \citep{2012RSPTA.370.2765A} limited to effective temperatures between 400 K and 2500 K with a step size of 100 K, surface gravities between 0.0 dex and 5.5 dex with a step size of 0.5 dex, and metallicity {[Fe/H]} = 0. We then integrated the flux of each model in the grid over the spectral response curves of each observed filter to find the scaling parameter that best matched the model to the SED of the companion, characterised as the value that minimizes the Euclidean norm of the residual vector between the two. The overall best fit model is then identified as the one that results in the minimum residual compared to the SED. In lieu of spectral response curves for the SPHERE IFS and GPI spectral data, we treated the spectral response of each wavelength channel as a Gaussian corresponding to the resolution of the spectrograph in the relevant band \citep{2017A&A...603A..57S}. When the fitting procedure described above was performed on the full spectrum of HR~2562~B, the minimum residual is given by a model with T\textsubscript{eff} = 1700 K and log(g) = 5.0 dex, shown alongside the SED as a purple line in Figure \ref{fig:model_multi}. As the MagAO and SPHERE photometry were obtained concurrently on the same night, we also performed the fitting procedure to this subset of the data. On the other hand, as the GPI data were not obtained concurrently with the MagAO data, we did not apply the fitting procedure to that subset of data. The best fit model to the subset of concurrent MagAO and SPHERE photometry alone instead has T\textsubscript{eff} = 1200 K and log(g) = 4.0 dex, shown as a green line. The reduced chi-square values of the fits to the full spectrum of HR~2562~B and to the MagAO + SPHERE-only subset of data are 4.40 and 3.86, respectively, suggesting that neither model is a particularly satisfying match for the corresponding data. Indeed, while the T\textsubscript{eff}~=~1700 K model is statistically the best fit to the full SED and is a closer match to the amplitude of the peaks in the GPI spectrum, it is almost flat in the K-band and visibly fails to capture the wide absorption bands seen in the SED of HR~2562~B. Conversely, while the T\textsubscript{eff}~=~1200 K model does show these absorption features, the amplitudes of the peaks miss those of the GPI spectrum. We attempt to explain these differences between the synthetic spectra and the observational data, and the corresponding absence of a strong best fit result, in Section \ref{discussion_models}. We assess the effect of the photometric measurement errors on the outcome of this fitting procedure by iterating 10$^{5}$ times, varying the data flux values across Gaussian distributions centered on the original value, where the uncertainty on the original value is used as the standard deviation of the sampling. This statistical error on the derived physical properties of the companion is given by the 2.5 and 97.5 percentiles of the corresponding distribution of models \citep{2020MNRAS.492..431B}. We then use the largest of either the statistical error or the BT-Settl model grid spacing of $\pm$100~K in temperature and $\pm$0.5~dex in surface gravity as our reported uncertainties on these physical parameters. By integrating over the full wavelength range of the models and accounting for the distance to the system, we further infer the companion luminosity in each case. The estimates provided by the procedure described above, considering the full SED and separately the MagAO + SPHERE-only subset of data, are given in Table \ref{table:model_posteriors}. The scaling parameter is equivalent to R$^2/$D$^2$, where R is companion radius and D is the distance to the system (where D is well constrained), so we are further able to infer radius estimates for each best fit case. The fit to the full SED yields a radius of R = 0.56$^{+0.02}_{-0.01}$~R\textsubscript{Jup}, whereas in the MagAO + SPHERE-only case we find R = 0.89$^{+0.14}_{-0.27}$~R\textsubscript{Jup}. The reported uncertainties on the luminosity and radius estimates are the statistical errors. These results and the differences between those derived in each fitting case are discussed further in Section \ref{discussion_models}, where we note the likely unphysical radius derived from the full SED.
\subsubsection{Empirical templates}\label{results_templates}
Noting the differences between synthetic spectra and the observations, we further performed the fitting procedure described in Section \ref{results_models} using empirical template spectra of field L and T dwarfs from the SpeX Prism Spectral Libraries \citep{2014ASInC..11....7B}. These templates are limited in wavelength range to 0.65 - 2.56 \textmu m, and so do not extend to the 3.94 \textmu m position of our MagAO datapoint for the required spectral types. Nonetheless, we proceeded with a comparison to these templates to further investigate the differences between fits to the SPHERE and GPI data, as well as to determine a spectral type for HR~2562~B.
We find the best fit template to the combined SPHERE and GPI data to be that of 2MASSW J2244316+204343 \citep{2003ApJ...596..561M, 2008ApJ...686..528L}, which has a spectral type of L7.5$\pm$2, plotted in Figure \ref{fig:template_multi} as a pink line. The same best fit template is obtained when the fitting procedure is performed for the GPI data alone, but fitting to the SPHERE data alone instead best matches the spectrum of SDSS J151643.01+305344.4 \citep[spectral type T0.5$\pm$1,][]{2006AJ....131.2722C, 2010ApJ...710.1142B}. This template is shown in Figure \ref{fig:template_multi} as a grey line. We therefore consider HR~2562~B to have a spectral type at the L/T transition, and discuss this interpretation further in \ref{discussion_templates}.
\section{Discussion}\label{discussion}
\subsection{Photometry}\label{phot}
In Section \ref{phot_measure}, we report a marginal detection of HR~2562~B with a S/N of 3.04 in the final image produced by cADI at a position which matches that measured by \citet{2018A&A...615A.177M} and \citet{2018A&A...612A..92M}, who observed this companion on the same night using SPHERE. However, this value is notably lower than the S/N reported by \citet{2018A&A...612A..92M}, who detected HR~2562~B at a S/N of $\sim$20 in their final SPHERE IRDIS image, and $\sim$30 in their final SPHERE IFS image. Although \citet{2016ApJ...829L...4K} do not provide the S/N of the detections of HR~2562~B in their final GPI images, it is clear that these are on a similar order to the SPHERE detections. This difference can primarily be explained by comparing the bandwidths of each set of observations. For our MagAO+vAPP observations, we used a 3.94~\textmu m narrow-band filter with a width of 90~nm. This is significantly narrower than the H broad-band SPHERE IRDIS filter, which has a width of 290~nm. and the wavelength ranges covered by the final SPHERE IFS and GPI images, which are composed of spectral datacubes collapsed across their respective wavebands. Our lower S/N is therefore unsurprising. The flux measurement error of our MagAO datapoint is comparable to those of the individual spectral datapoints of SPHERE IFS and GPI. The use of a broad-band filter may be preferable if one were to conduct a blind search for undiscovered companions, where the position is not already known, as the wider wavelength coverage will enable the capture of greater companion flux and hence a stronger initial detection. However, the polarization grating of the MagAO vAPP causes wavelength-dependent smearing of the PSFs across the detector when broad-band filters are used. An additional processing step is therefore required to either extract the resulting low-resolution spectra or recombine the PSFs along the axis of the vAPP. Alternatively, broad wavelength coverage can be achieved without lateral smearing by using a vAPP in combination with an integral field spectrograph, or a vAPP with a 360\textdegree{} dark hole, which is not affected by such smearing as a second polarization grating is used to recombine the beams on axis \citep{2020PASP..132d5002D}. Another factor affecting the strength of our companion recovery is the thermal background flux arising from both the sky and the instrumentation itself, which is far greater at 3.94~\textmu m than at the shorter wavelengths used to observe HR~2562 in previous studies \citep{2000PASP..112..264L}. The difference in the size of the telescopes used in these observations further contributes to the lower S/N reported in this work; the 6.5-m Magellan Clay Telescope used for these observations is slightly smaller than the 8.1-m Gemini South telescope, on which GPI is installed, and the 8.2-m VLT Unit Telescope, where SPHERE is installed. Lastly, the combination of field-stabilised and pupil-stabilised observations composing this dataset may also have had some impact on the S/N, as the field-stabilised parts may contribute some companion signal to the reference PSF removed by cADI.

\begin{figure}
	\includegraphics[scale=0.5]{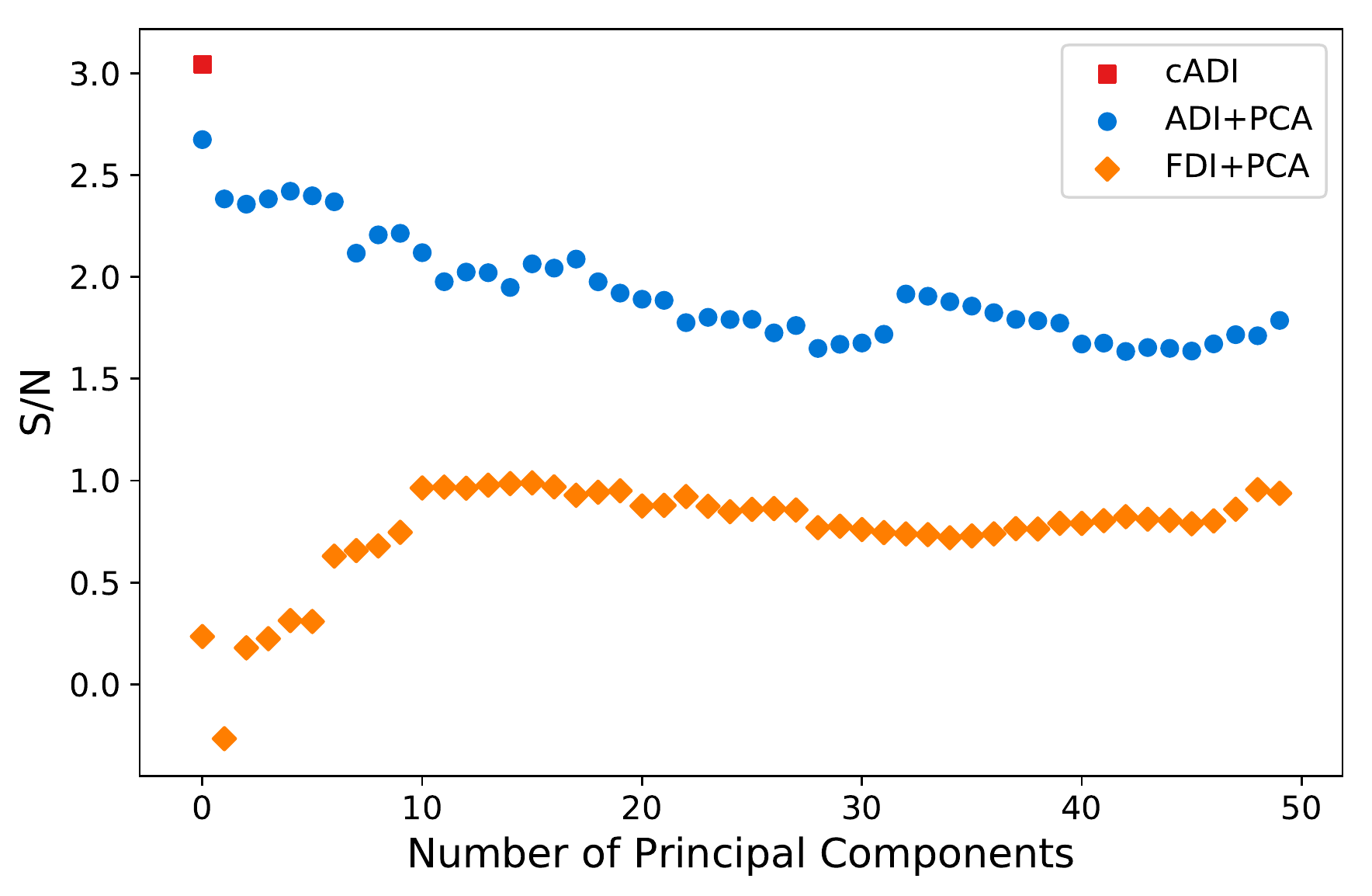}
    \caption{The S/N at the companion location in the final images produced by each algorithm, as a function of the number of principal components removed in each case. Although ADI+PCA is more effective at removing noise than cADI (see Figure \ref{fig:final_images}), its performance is negatively impacted by oversubtraction which reduces the signal of the companion. As HR~2562~B is not detected in the FDI+PCA images, the variation of the noise is greater than the peak flux at the companion location, leading to a S/N smaller than one. Removing additonal principal components has the effect of increasingly homogenising the image, causing the S/N at the companion location to tend towards one.}
    \label{fig:sn_components}
\end{figure}

Due to the small angular coverage of the dark holes, residual noise structure from the vAPP PSFs in the contaminated regions, and the non-standard combination of field and pupil-stabilised observations comprising this dataset, it is not possible to produce a meaningful assessment of the detection limits reached by each algorithm in this particular case. In the final images, not enough space remains to place the number of photometric apertures required to validly estimate the noise term, especially at small separations \citep{2018AJ....155...19J}. Furthermore, these detection limits will vary significantly not only with angular separation from HR~2562, but also depending on the position angle being considered. In lieu of such measurements of the detection limits, we include an alternative, if limited, comparison of the performance of the three algorithms applied to this data. Figure \ref{fig:sn_components} shows the S/N at the location of HR~2562~B in the final images, as produced by each algorithm, as a function of the number of principal components removed in each case. As stated in Section \ref{postprocess}, cADI produces the image with the highest S/N recovery of HR~2562~B (S/N=3.04). Although ADI+PCA is far more effective than cADI at reducing noise, even succeeding in removing the residual contamination from the vAPP PSF between the dark holes (dotted segments, Figure \ref{fig:final_images}), its performance is limited by oversubtraction which reduces the signal of the companion. This can again be attributed to the non-standard combination of field and pupil-stabilised data, due to which the companion is fixed in the same location for a significant fraction of the observing sequence. It is unsurprising that some degree of companion self-subtraction occurs when our data is processed with ADI+PCA as the first component of this algorithm is simply the mean combination of the input images orthogonalised with respect to the PCA basis. While this effect likely also impacts the signal of the cADI detection, the reference PSF in this case is constructed using a median combination of the data, which will capture a lesser degree of companion flux when the majority of the observations are pupil-stabilised. Removing additional PCA components gradually suppresses the companion signal further, increasingly homogenising the image. In the case of FDI+PCA, with which the companion is not detected, the variation of the noise is greater than the peak flux at the companion location when a small number of principal components are applied, leading to a S/N smaller than one. As with PCA, this variation is gradually suppressed with additional components.

\subsection{Flipped Differential Imaging (FDI)}\label{fdi}
\begin{figure}
	\includegraphics[scale=0.5]{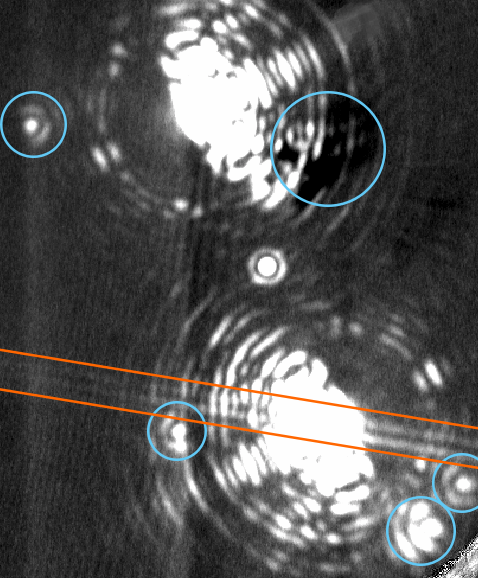}
    \caption{A background-subtracted and median combined frame from the 3.94~\textmu m MagAO+vAPP observations of HR~2562, cropped around the vAPP PSFs. A number of PSF symmetry-breaking artefacts are visible, including reflection ghosts (highlighted in blue) and a bright spike of scattered light that passes directly through the dark hole of the bottom PSF (in orange). The frame is not aligned to north.}
    \label{fig:ghosts}
\end{figure}

Although HR~2562~B is visible in the final cADI- and ADI+PCA-processed images, we are unable to detect it in the image resulting from the PCA-based FDI procedure. As FDI+PCA is inherently reliant on the symmetry of the PSFs along the axis of the vAPP (and by extension, the response of the detector to incoming flux), artefacts such as reflection ghosts can have a significant effect on the ability of the algorithm to achieve optimal flux suppression in the vAPP dark holes \citep{2017ApJ...834..175O}. \citet{2018SPIE10703E..2TL} characterised many such artefacts on the Clio2 camera, including some that are only visible following a background subtraction, and several that scale with increased incoming flux, such as amplifier crosstalk \citep{2015ApJ...815..108M}. A number of these effects and their impact on the vAPP dark holes can be seen in Figure \ref{fig:ghosts}. In particular, a bright spike of scattered light passes directly through the dark hole of the bottom coronagraphic PSF while the top remains unaffected. Furthermore, this artefact does not appear in the same way when the vAPP is positioned in the alternate nod position. The symmetry of the coronagraphic PSFs was likely further impacted by the wind-driven halo effect described by \citet{2018A&A...620L..10C, 2020A&A...638A..98C} and \citet{2018SPIE10703E..6EM, 2019JATIS...5d9003M}, which results when atmospheric turbulence above the telescope pupil, primarily in the jet stream layer, varies at a rate faster than can be corrected for by the deformable mirror of the adaptive optics system. Indeed, the characteristic `butterfly pattern' of wind-driven halo can be seen in the final FDI+PCA image of Figure \ref{fig:final_images} as the extended bright and dark patches either side of the masked inner region. Even if the butterfly pattern were perfectly aligned along the axis of the vAPP, interference between scintillation effects and the lag in adaptive optics correction gives rise to an asymmetry in the butterfly pattern itself. This asymmetry is wavelength-dependent, growing stronger at longer wavelengths. As these instrumentational and atmospheric effects all negatively impact the symmetry between the two coronagraphic stellar PSFs, it is likely that the reference PSF constructed using FDI+PCA on our HR~2562 data was a poor match for the opposing coronagraphic stellar PSF, thus explaining the non-detection of the companion in the final image. Companion detection using the first night of observations was further inhibited by the increased effect of readout noise resulting from the shorter exposure time. Although successful photometric extraction via FDI+PCA was not possible within the limitations of the data presented here, it could be a potentially effective strategy for future observations if a high enough degree of symmetric precision can be reached between the two coronagraphic PSFs of the vAPP. FDI+PCA is built on the approach of \citet{2017ApJ...834..175O}, who use the opposing vAPP coronagraphic PSF as a reference directly, without PCA. When applied to MagAO+vAPP observations obtained under excellent atmospheric conditions, they find that this technique reaches contrasts up to 1.46 magnitudes deeper than cADI. They further cite the case of \citet{2015ApJ...802...12D}, who apply ADI+PCA to a non-coronagraphic PSF under 180\textdegree{} rotation to create a reference PSF, and achieve an order of magnitude improvement in contrast at small separations (compared to when the Locally Optimised Combination of Images algorithm, LOCI, is applied to ADI data, \citealt{2007ApJ...660..770L}). Considering these results, \citet{2017ApJ...834..175O} conclude that a PCA-based algorithm such as FDI+PCA should produce an improved reference PSF and achieve even deeper contrasts compared to when the opposing vAPP PSF is used as a reference without PCA. However, as the observations here are not fully optimised for high-contrast imaging, and further contain the symmetry-breaking artefacts described above, they serve to highlight where this technique can break down. An analysis using better optimised data is required to fully determine the potential of FDI+PCA and to compare its performance to that of other post-processing algorithms. Coronagraphic simulations could further be used to assess the extent to which different symmetry-breaking factors limit the performance of FDI+PCA and establish mitigation strategies for the most significant contributors. Although instrumental artefacts such as reflection ghosts may be challenging to remove completely, asymmetries arising from effects such as wind-driven halo vary between observations, and will be increasingly manageable with ongoing advancements in wavefront sensing and predictive control \citep{2017arXiv170700570G, 2018SPIE10703E..1TM, 2021A&A...646A.145M, 2018SPIE10703E..1UJ, 2019A&A...632A..48B, 2020A&A...636A..81V}. A number of 180\textdegree{} coronagraphs are currently installed on instruments at other telescopes, including SCExAO/CHARIS on the 8.2-m Subaru Telescope \citep{2017SPIE10400E..0UD} and LMIRcam/ALES on the 8.4-m Large Binocular Telescope \citep{2014OExpr..2230287O}, and several are planned for future instruments, such as MagAO-X on Magellan \citep{2019JATIS...5d9004M}, ERIS on the VLT \citep{2018SPIE10702E..3YB, 2018SPIE10702E..46K}, and METIS on the ELT \citep{2016SPIE.9909E..73C, 2018SPIE10702E..1UB}. A thorough evaluation and comparison of the different post-processing algorithms that can be applied to vAPP data is essential if observations using vAPP coronagraphs are to be used to their full potential. While the effectiveness of FDI+PCA has not yet been demonstrated, it is an alternate processing pathway uniquely available to the vAPP and thus could prove advantageous if the limiting factors can be overcome.
\subsection{Companion characterisation}\label{parameter_analysis}
\subsubsection{Theoretical atmospheric models}\label{discussion_models}
The fitting of BT-Settl atmospheric models to the full SED and separately to the MagAO + SPHERE-only subset of data produces substantially different physical parameters for HR~2562~B (see Table \ref{table:model_posteriors}). Our values for the concurrent MagAO + SPHERE data are in good agreement with \citet{2018A&A...612A..92M}, who found T\textsubscript{eff} = 1100$\pm$200~K and log(g)~=~4.75$\pm$0.41~dex by fitting several atmospheric models to the SPHERE data only, including the BT-Settl models used in this work. \citet{2016ApJ...829L...4K}, whose analysis of the GPI spectra by way of evolutionary models produces T\textsubscript{eff}~=~1200$\pm$100~K and log(g)~=~4.7$\pm$0.32~dex, is also in good agreement. Our calculated radius from the MagAO + SPHERE-only case is R = 0.89$^{+0.14}_{-0.27}$~R\textsubscript{Jup}, which is consistent within 1$\sigma$ to \citet{2016ApJ...829L...4K}, who estimated a radius of R = 1.11$\pm$0.11 R\textsubscript{Jup} using the evolutionary models from \citet{2008ApJ...689.1327S}. However, the temperature and surface gravity values produced by fitting the full SED with BT-Settl are notably higher, and T\textsubscript{eff}~=~1698$\pm$100~K is inconsistent with the literature. Furthermore, the sub-Jupiter value for the radius derived from this analysis (R = 0.56$^{+0.02}_{-0.01}$~R\textsubscript{Jup}) is unphysically small due to the pressure of degenerate electrons in the interior of brown dwarfs \citep{2009AIPC.1094..102C}. We also note that neither of the best fit models resulting from our analysis is a strongly compelling match for the SED of the companion when inspected visually. The T\textsubscript{eff}~=~1700 BT-Settl model, although statistically the best fit to the full SED, does not feature the wide absorption bands visible in the companion SED. While these bands are seen in the T\textsubscript{eff}~=~1200 BT-Settl model, this model instead fails to match the absolute fluxes of the GPI observations. In both cases, the reduced chi-square values of the fits suggest that a good fit is not achieved, although arguably one might favour the models that produce physically plausbile radii. Such wide-ranging best fit parameters and low radii estimates resulting from fits of atmospheric grid models to observations of substellar objects with L/T spectral types have been reported previously, with an apparent dependence on both wavelength range and the specific wavebands included in the fit, as well as the models used \citep{2021AJ....161....5W, 2020AJ....160..262S, 2020AJ....160..207W, 2017AJ....154...10R, 2015ApJ...815..108M}. \citet{2014A&A...564A..55M} found that although the BT-Settl models are largely successful at reproducing the SEDs of L-type objects, they do not always match the redness of the spectral slope in the near-infrared, suggesting that the cloud models do not include enough dust at high altitudes. Indeed, despite accounting for non-equilibrium chemistry and aiming to reproduce the L/T transition in brown dwarfs, \citet{2016A&A...587A..58B} demonstrated that the BT-Settl models can struggle to simultaneously produce good matches for both the shape and absolute fluxes of the SEDs of the highly red HR8799 planets, leading to underestimated radii \citep{2008Sci...322.1348M, 2012ApJ...754..135M}. The challenge in fitting these models to the SED of HR~2562~B (and the resulting wide range of physical parameters) could therefore be due to the slightly enhanced flux in the K-band compared to the J- and H-bands, potentially caused by the presence of dust in the high altitude cloud layer. Although HR~2562~B is not so strongly red as HD~206893~B (the reddest substellar object observed to date, and a system with remarkably similar architecture to HR~2562 \citep{2017A&A...597L...2M, 2021AJ....161....5W}, \citet{2018A&A...612A..92M} show that it is slightly redder than other objects at the L/T transition, such as HN~Peg~B (which is of comparable mass and age \citep{2007ApJ...654..570L}). We also consider the possibility that the model fit to the full SED could be impacted by systematic differences between the SPHERE and GPI photometry. Although the SPHERE IFS and GPI spectra are comparable where they overlap in the J-band, it could be argued that there is a small difference between the two, due to differences in the flux calibration or otherwise. However, a constant offset applied to bring the two level would still fail to bring the GPI K-band data to match the best fit models in either case. Brown dwarfs are known to vary in time, and that such variability can manifest differently at different wavelengths \citep{2013ApJ...778L..10B, 2017ApJ...842...78V, 2018MNRAS.474.1041V, 2018AJ....155...11M, 2020ApJ...893L..30B}. This could influence the shape and absolute fluxes of the SED of HR~2562~B, including any difference between the SPHERE and GPI photometry, although the SPHERE and MagAO+vAPP data are concurrent. A large Spitzer survey of isolated brown dwarfs concluded that photometric variability is ubiquitous for L and T dwarfs, with some exhibiting up to $\sim$5\% amplitude variations \citep{2015ApJ...799..154M}. Recent studies have provided further evidence that brown dwarfs close to the L/T transition present the most variability, attributing the variations to patchy clouds (or clouds of varying thickness) rotating in and out of view throughout the rotation periods of the objects \citep{2016ApJ...825...90K, 2018ApJ...854..172C, 2019MNRAS.483..480V, 2020AJ....159..140Z}. The 3.94 \textmu m MagAO+vAPP measurement matches the T\textsubscript{eff} = 1200 K, log(g) = 4.0 dex best fit model to the MagAO + SPHERE data, but the error bar spans a wide range of BT-Settl models with different physical parameters, including the T\textsubscript{eff} = 1700 K, log(g) = 5.0 dex best fit model to the full SED. Although this datapoint alone is therefore unable to further constrain the physical parameters of HR~2562~B, we can conclude that its flux at this wavelength is not unusual for an object of the range of temperatures and surface gravities previously derived for HR~2562~B in the literature and lend additional weight to these values. It is clear from Figure \ref{fig:model_multi} that complementary observations in the 2.4-3.2 \textmu m region would be most effective in distinguishing models due to the onset of significant absorption bands in this region for cooler objects. To overcome telluric bands in this window, this will likely require space-based instruments such as the James Webb Space Telescope (JWST, \citealt{2006SSRv..123..485G, 2018SPIE10698E..09P}), or ground-based high resolution spectroscopy \citep{2013MNRAS.436L..35B, 2014Natur.509...63S, 2016A&A...593A..74S, 2018A&A...617A.144H}. JWST/MIRI will further provide charaterisation at wavelengths longer than $\sim$5 \textmu m, with observations of HR~2562~B already planned as part of Cycle 1 GTO Program 1241 (PI:~M.~Ressler).
\subsubsection{Empirical templates}\label{discussion_templates}
The fitting of empirical template spectra to the SPHERE + GPI data together gave a best fit object with a spectral type of L7.5$\pm$2, while the best fit to the SPHERE data alone was an object with a spectral type of T0.5$\pm$1, suggesting that HR~2562~B has a spectral type within the L/T transition regime. These results are consistent with those previously reported. For example, \citet{2018A&A...612A..92M} compared their extracted spectrum to a range of template spectra between L5 and T5.5 and concluded that an early T (T2-T3) spectral type was the best match overall, but that their SPHERE IRDIS H broad-band datapoint was better described by a late L spectra. Similarly, \citet{2016ApJ...829L...4K} found that the GPI SED in full is not matched perfectly by the empirical spectra of any other object but that objects with spectral types between L3.5 and T2 do offer good fits to individual wavebands, concluding a spectral type of L7$\pm$3 while noting that brown dwarfs can have very different colours while possessing similar spectral features \citep{2003csss...12..120L, 2018AJ....155...34C}. This also reflects the issue described in Section \ref{discussion_models}, where fitting atmospheric models to different wavelength ranges or individual wavebands can produce different results. One might further consider that brown dwarf companions and field brown dwarfs could have different properties, and that the spectra of field brown dwarfs may therefore not be the ideal comparison to those of bound substellar companions. While \citet{2016ApJ...833...96L} found evidence that young brown dwarf companions with late-M and L spectral types may form distinct sequences on infrared colour-magnitude diagrams compared to the field dwarf population, their analysis suggests that the two populations are broadly consistent in the L/T transition regime (noting however, that the L/T transition lies beyond the spectral type and colour range of their fits). \citet{2018A&A...612A..92M} stated that observations on a wider wavelength range would be needed to completely disentangle the spectral classification of HR~2562~B. While the 3.94 \textmu m MagAO datapoint can potentially assist with this, there remains a lack of L and T dwarf empirical template spectra in the literature that cover the wavelength range up to and including 3.94 \textmu m. Without such benchmark spectra for comparison, attaining a model-independent classification of the spectral type of HR~2562~B remains a challenge.
\subsubsection{Mass estimation}\label{mass_estimation}
To derive a range of possible values for the mass of the companion, we evaluated our inferred luminosities with BT-Settl \citep{2012RSPTA.370.2765A, 2015A&A...577A..42B} and AMES-Dusty \citep{2001ApJ...556..357A, 2000ApJ...542..464C} isochrones across the system age range of 450$^{+300}_{-250}$ Myr range found by \citet{2018A&A...612A..92M}. Although this process could also be performed using our derived values for effective temperature or surface gravity, luminosity is generally much less model dependent \citep{2016A&A...587A..58B}. The two different sets of models account for atmospheric dust formation in different ways; the BT-Settl models do so by way of a parameter-free cloud model whereas the AMES-Dusty models assume that dust is formed in equilibrium with the gas phase. The results of this mass evaluation are presented in Table \ref{table:model_posteriors}, alongside the corresponding values of mass ratio with respect to the primary, $q$. Considering the spread of these results, we report a weighted average value of 29$\pm$15 M\textsubscript{Jup} as our final mass estimate with a corresponding mass ratio $q$ of 0.020$\pm$0.011. This is consistent with the range of values found by \citet{2018A&A...612A..92M} by comparing evolutionary models to the SPHERE photometry in each band individually using the same age range, as well as their final reported value of 32$\pm$14 M\textsubscript{Jup}. A similar estimate of 30$\pm$15 M\textsubscript{Jup} was found by \citet{2016ApJ...829L...4K}, who assumed a slightly higher and wider age range of 300-900 Myr. As previously noted by \citet{2018A&A...612A..92M}, these values are consistent with those of a brown dwarf with a late-L/early-T spectral type when compared to the dynamical mass measurements of ultracool M7-T5 objects by \citet{2017ApJS..231...15D}, matching the spectral classification in Section \ref{discussion_templates}. The wide uncertainties on these estimates are dominated by the uncertainty on the age of the system, which is not well constrained for HR~2562, and reflect the strong dependence of substellar companion mass measurements on system age. Either a dynamical mass measurement or improved constraints on the age of the system are therefore crucial if the mass of HR~2562~B is to be constrained further.
\section{Conclusions}\label{conclusions}
We present a S/N=3.04 recovery and tentative characterisation of a companion in the lesser studied L-band regime using a vector Apodizing Phase Plate coronagraph in observations obtained with MagAO+vAPP, recovering the known brown dwarf companion to HR~2562 previously studied with GPI \citep{2016ApJ...829L...4K} and concurrently with SPHERE \citep{2018A&A...612A..92M, 2018A&A...615A.177M}. We processed our 3.94~\textmu m images using cADI, ADI+PCA, and a newly-developed algorithm, FDI+PCA. We measure the companion 3.94~\textmu m contrast to be ($3.05\pm1.00$)~$\times$~10$^{-4}$ relative to the host star, which is equivalent to a physical flux of ($1.3\pm0.4$) $\times$  10$^{-13}$ erg s$^{-1}$ cm$^{-2}$ micron$^{-1}$. The companion is visible in images produced by applying cADI and ADI+PCA to the observations from the second night. The highest S/N (= 3.04) is produced by cADI. Although this S/N is low, the companion recovery is further supported by its position, which matches that measured by \citet{2018A&A...615A.177M} in observations obtained on the same night. This S/N is lower than those of literature detections of HR~2562~B, but this can primarily be attributed to the significantly narrower filter used in this work and the higher thermal background at 3.94~\textmu m. We do not detect HR~2562~B in the final images produced from the first night of observations, which did not cover sufficient field rotation to prevent self-subtraction when applying post-processing algorithms. Performing observations in pupil-stabilised mode, with the field of view rotating, is therefore likely necessary to detect high-contrast systems like HR~2562~B with this instrument setup. We describe FDI+PCA, a new post-processing algorithm that uses the symmetry of the vAPP PSFs to construct a reference PSF for subtraction from the data, removing quasistatic speckle noise. Although we were unable to recover the companion in our FDI+PCA processed image, we explain the impact of instrumental scattered light and wind-driven halo which degrade the symmetry of the vAPP and consequently reduce the effectiveness of the algorithm. FDI+PCA may still prove effective for future datasets that use a 180\textdegree{} vAPP, obtained under more optimal atmospheric conditions or on instruments with fewer scattered light artefacts, but further analysis is required to assess its potential. Broad-band filters may be preferred for MagAO+vAPP observations conducting blind searches for undiscovered companions as wider wavelength coverage will enable stronger detections, despite the lateral smearing of the PSFs that occurs when such filters are used. This wavelength dependent smearing can be handled through additional processing to either extract the resulting low-resolution spectra or collapse the PSFs along the axis of the vAPP. This wavelength-dependent smearing can alternatively be avoided by using a 360\textdegree{} vAPP coronagraph, which does not have such smearing even when broad-band filters are used \citep{2020PASP..132d5002D}. Wide wavelength coverage can also be achieved when vAPPs are combined with integral field spectrographs \citep{2014OExpr..2230287O}. Nonetheless, MagAO+vAPP still allowed a measurement in the lesser studied L-band regime. We fit BT-Settl atmospheric models to our 3.94 \textmu m flux in combination with literature spectral data from GPI \citep{2016ApJ...829L...4K} and SPHERE \citep{2018A&A...612A..92M}, and find different results depending on the wavebands included in the fit. We do not find a single model that is a convincing match to the SED, and instead find a wide range of allowable values, including 1200$\leq$T\textsubscript{eff}(K)$\leq$1700 and 4.0$\leq$log(g)(dex)$\leq$5.0 for the companion; dependent on which wavelength regions are fitted. Although we were therefore unable to significantly further constrain the physical parameters of the companion, the consistent measurements lend additional weight to those derived in the literature and highlight the degeneracies that arise from fitting atmospheric models to brown dwarf atmospheres. Complementary observations at 2.4-3.2 \textmu m will help distinguish cooler brown dwarfs due to the onset of absorption bands at this wavelength region. Comparing the SED of the companion to empirical template spectra, we conclude that HR~2562~B has a spectral type at the L/T transition. However, the unavailability of templates with 3.94~\textmu m coverage precluded us from including our MagAO datapoint in this fit. We also evaluate the inferred luminosities using BT-Settl and AMES-Dusty isochrones across the system age range of 450$^{+300}_{-250}$ Myr, deriving a mass estimate for HR~2562~B of 29$\pm$15 M\textsubscript{Jup}, in good agreement with the values found by \citep{2016ApJ...829L...4K} and \citep{2018A&A...612A..92M} and consistent with the mass of a late-L/early-T type brown dwarf. As companion mass is highly dependent on system age, either a precise dynamical mass measurement or improved constraints on the age of the system are crucial if the mass of HR~2562~B is to be constrained further.

\section*{Acknowledgements}\label{ack}
The authors thank Quinn Konopacky and Dino Mesa for sharing the GPI and SPHERE IRDIS + IFS\footnote{Based on observations collected at the European Southern Observatory under ESO programme(s) 198.C-0209(D).} spectra of HR~2562~B, respectively. The authors would also like to thank Gilles Otten, Jos de Boer, Sebastiaan Haffert, and Steven Bos for valuable discussions that improved this work. We thank our anonymous referee whose comments helped to improve this manuscript, and for their timely response in these unusual times. We are especially grateful to France Allard for providing the community access to the PHOENIX and BT-Settl model database for so many years. It has been enormously impactful and we hope that her legacy continues long in her absence.

BS is fully supported by the Netherlands Research School for Astronomy (NOVA). JLB acknowledges funding from the European Research Council (ERC) under the European Union’s Horizon 2020 research and innovation program under grant agreement No 805445. This work was performed in part under contract with the Jet Propulsion Laboratory (JPL) funded by NASA through the Sagan Fellowship Program executed by the NASA Exoplanet Science Institute. KMM's work is supported by the NASA Exoplanets Research Program (XRP) by cooperative agreement NNX16AD44G. The research of DD and FS leading to these results has received funding from the European Research Council under ERC Starting Grant agreement 678194 (FALCONER). This paper includes data gathered with the 6.5 meter Magellan Telescopes located at Las Campanas Observatory, Chile. This paper uses observations obtained at the Gemini Observatory, which is operated by the Association of Universities for Research in Astronomy, Inc., under a cooperative agreement with the NSF on behalf of the Gemini partnership: the National Science Foundation (United States), the National Research Council (Canada), CONICYT (Chile), the Australian Research Council (Australia), Ministério Ciência, Tecnologia e Inovação (Brazil) and Ministerio de Ciencia, Tecnología e Innovación Productiva (Argentina).
This publication makes use of VOSA, developed under the Spanish Virtual Observatory project supported by the Spanish MINECO through grant AyA2017-84089. VOSA has been partially updated by using funding from the European Union's Horizon 2020 Research and Innovation Programme, under Grant Agreement \textnumero 776403 (EXOPLANETS-A). This publication makes use of data products from the Two Micron All Sky Survey, which is a joint project of the University of Massachusetts and the Infrared Processing and Analysis Center/California Institute of Technology, funded by the National Aeronautics and Space Administration and the National Science Foundation. This publication makes use of data products from the Wide-field Infrared Survey Explorer, which is a joint project of the University of California, Los Angeles, and the Jet Propulsion Laboratory/California Institute of Technology, funded by the National Aeronautics and Space Administration. This work has made use of data from the European Space Agency (ESA) mission
{\it Gaia} (\url{https://www.cosmos.esa.int/gaia}), processed by the {\it Gaia} Data Processing and Analysis Consortium (DPAC, \url{https://www.cosmos.esa.int/web/gaia/dpac/consortium}). Funding for the DPAC has been provided by national institutions, in particular the institutions participating in the {\it Gaia} Multilateral Agreement. This research has benefitted from the SpeX Prism Spectral Libraries, maintained by Adam Burgasser at \url{http://pono.ucsd.edu/~adam/browndwarfs/spexprism}. This research has made use of NASA’s Astrophysics Data System. This research has made use of the SIMBAD database, operated at CDS, Strasbourg, France \citep{2000A&AS..143....9W}. This research made use of ds9, a tool for data visualization supported by the Chandra X-ray Science Center (CXC) and the High Energy Astrophysics Science Archive Center (HEASARC) with support from the JWST Mission office at the Space Telescope Science Institute for 3D visualization \citep{2003ASPC..295..489J}. This work makes use of the Python programming language\footnote{Python Software Foundation; \url{https://www.python.org/}}, in particular packages including NumPy \citep{oliphant2006guide, van2011numpy}, Astropy \citep{astropy:2013, astropy:2018}, SciPy \citep{2020SciPy-NMeth}, scikit-image \citep{scikit-image}, Photutils \citep{larry_bradley_2017_1039309}, and Matplotlib \citep{Hunter:2007}.
\section*{Data Availability}\label{ava}
The data from the MagAO observations underlying this article are available in the Research Data Management Zenodo repository of the Anton Pannekoek Institute for Astronomy, at \url{https://doi.org/10.5281/zenodo.4333200}.





\bibliographystyle{mnras}
\bibliography{bibliography} 









\bsp	
\label{lastpage}
\end{document}